\documentstyle{article}
\title{\LARGE What our world might be like 
(as alternative to Minkovski space-time with Poincare group of motion)}
\author{A.~N.~Leznov\thanks{ Universidad Autonoma del Estado de Morelos, CCICAp,Cuernavaca, Mexico}} \date{}

\newcommand{\rig}[2]{\stackrel{#2\rightarrow}{#1}}

\begin{document}
\maketitle

\maketitle

\begin{abstract}
It is proposed four dimensional curved space-time with de-Sitter group of motion. Theory contain free dimension constants of length, impulse and action. Under infinite values of these parameters theory pass to usual Minkowski space-time with Poincare group of motion. In modified space gauge invariant field theory is constructed.    
\end{abstract}

\section{Introduction}

Imagine for a moment that theory of relativity was discovered by the following logical pure theoretical manner. The physic of Newton is invariant (except of translations) to rotations 
$\rig{l}{}_i=\epsilon_{ijk} x_jp_k$ and transformation of Galileo $\rig{g}{}_i=tp_i$. These generators satisfy the following commutation relations
$$
[l_i,l_j]=\epsilon_{ijk}l_k, \quad [l_i,g_j]=\epsilon_{ijk}g_k, \quad [g_i,g_j]=0
$$
If one will try to modify this relation the obvious and simplest way is to add to the left side of the last equation term proportional to ${1\over c^2}\epsilon_{ijk}l_k$. With dimension of $c$ as velocity. Indeed $\rig{l}{}$ is dimensionless and $\rig{g}{}$ have inverse of velocity dimension. After this it would be possible to construct theory invariant to new group of motion which will be different from mechanics of Newton only in the domain of velocity near to $c$.

The same logic was used in attempts to construct quantum spaces.  But preliminary the goal was change the properties of Minkowski space only on microscopic space-time distances. But from analyzes of dimensions it follows that all additional parameters arises in denominators exactly as $c$ in example above and to have correct results at least at distances of the solar system it is necessary to assume their values on the cosmical level.   

The most general form of the commutation relations between the elements of the quantum  four-dimensional space-time and its group of motion ($x$- coordinates, $p$- impulses, $F$- generators of Lorenz transformation, $I$ -unity element) are the following ones \cite{KHL} ( these relations are written under assuming that Lorenz group is included into the main laws of the nature)
$$
[p_i,x_j]=ih(g_{ij}I+{F_{ij}\over H}),\quad  [p_i,p_j]={ih\over L^2}F_{ij},
\quad [x_i,x_j]={ih\over M^2}F_{ij},
$$
\begin{equation}
[I,p_i]=ih({p_i\over H}-{x_i\over L^2}), \quad [I,x_i]=ih({p_i\over M^2}-
{x_i\over H}),\quad [I,F_{ij}]=0 \label{2}
\end{equation}
$$
[F_{ij},x_s]=ih(g_{is}x_j-g_{js}x_i),\quad [F_{ij},p_s]=ih(g_{is}p_j-g_{js}p_i)
$$
$$
[F_{ij},F_{sk}]=ih(g_{js}F_{ik}-g_{is}F_{jk}-g_{jk}F_{is}+g_{ik}F_{js})
$$
Commutation relations (\ref{2}) must be supported by some additional conditions which responsible for correct limit to usual Minkovski space-time in the infinite limit of dimensional parameters. Such conditions looks as
$$
IF_{i,j}={x_j p_i-x_i p_j+p_i x_j-p_j x_i \over 2}
$$
which in Minkovski limit ($I\to 1$) represent relation between angular moments and linear coordinates and impulses.
The additional conditions above allow to choose definite representation of algebra 
(\ref{2}) in the unique way \cite{LEZM}.

In the case of consideration problems in classical physics in (\ref{2}) it is necessary
to do transformation
$$ 
\lim {[A,B]\over ih}\to \{A,B\}
$$
(change all comutator of quantum theory on corresponding Poisson brackets) and consider 
(\ref{2}) on the level of functional algebra \cite{ESEN}).

Commutation relations of the quantum space contain 3 dimensional parameters
of the dimension length $L$, the impulse $Mc\to M$ and the action $H$. The
equalities of Jacobi are satisfied for (\ref{2}). It should be stressed the signs 
of $L^2,M^2$ are not required to be positive.
The limiting procedure $ M^2,H\to \infty$ leads to the space of
constant curvature, considered in connection with Column problem by
E.Schredinger \cite{SCH}, $L^2,H\to \infty$ leads to quantum space of Snyder
\cite{I}, $H\to \infty$ leads to Yangs quantum space \cite{Y}. Except of
$L^2,M^2$ parameter dimension of action $H$ was introduced in  \cite{KHL}.

The term quantum  space is not very adequate to this problem, because the modified classical dynamics may be considered in it (also as electrodynamics, gravity theory and so on). More correct but  not so short this problem may be called as possible generalized manifold of the world and its group of motion.  

In general (\ref{2}) is commutation relations one of real forms of six-dimensional rotation group. In the case $L^2\to \infty$ as it follows directly from (\ref{2}) commutation relations between $p_i,F_{i,j}$ exactly coincide with commutation relation of Poincare algebra and thus the last is the group of motion of the space-time manifold under consideration in this case.
Such version of quantum space was considered in \cite{LEZp}. We advice to the reader consider at first this more simple case. After this the reading of this paper will be more easier and understandable.

\section{Realization of quantum space on the base the usual Minkovski one}

The form of realization quantum space is demand for its understanding some knowledges from representation theory of semi-simple algebras \cite{LEZM}. Below we present some other form of realization (of course equivalent to previous one) but for checking of its validity sufficient knowledges of first chapters of usual curse of quantum mechanic.
Let us seek operators of four dimensional coordinates and impulses in a form
\begin{equation}
\bar p_i=a p_i+b A_i,\quad \bar x_i=c p_i+d A_i,\quad A_i=\rho x_i+[{x^2+1\over 2}p_i-x_i(x p)]
\label{DEF}
\end{equation}
where $\bar x,\bar p$ coordinates and impulses of real space-time and $p,x$ non
physical the same of Mikowski space with commutation relations $[p_i,x_j]=g_{i,j}$ and 
$x^2=x_0^2-(\rig{x}{})^2,(x p)=x_0 p_0-(\rig{x}{}\rig{p}{})$, $a,b,c,d,\rho$
arbitrary parameters which will be defined below. From definitions above the following commutation relations take place
$$
[A_i,A_j]=x_i p_j-x_j p_i\equiv F_{i,j},\quad [p_i,A_j]=g_{i,j}(\rho-(x p))+F_{i,j},\quad [(x p),x_i]=x_i,
$$
$$
[(x p),p_i]=-p_i,\quad [(x p),A_i]=A_i-p_i
$$ 
Now
$$
[\bar p_i,\bar x_i]=(ad-bc)g_{i,j}(\rho-(x p))+(ad+bc+bd)F_{i,j}=g_{i,j}hI+{h\over H}
\bar F_{i,j}
$$
From which we obtain relation and equation
$$
I={ad-bc\over h}(\rho-(x p))\equiv \delta (\rho-(x p)),\quad \bar F_{i,j}=\theta F_{i,j},\quad  ad+bc+bd={h\over H}\theta
$$
Absolutely by the same way from commutation relations between $\bar p_i$ and 
$\bar x_i$ we obtain additional pairs of equations
$$
b^2+2ab={h\theta\over L^2},\quad d^2+2dc={h\theta\over M^2}
$$
At last 
$$
[I,\bar x_i]=\delta [(\rho-(x p)),\bar x_i]=\delta(c+d)p_i-\delta d A_i=
{h\over M^2}\bar p_i-{h\over H}\bar x_i
$$
from what follows system of equations
$$
\delta(c+d)={ha\over M^2}-{hc\over H},\quad -\delta d={hb\over M^2}-{hd\over H}
$$
From $[I,\bar p_i]={h\over H}\bar p_i-{h\over L^2}\bar x_i$ absolutely by the same way it arises the system of equations
$$
\delta(a+b)={ha\over H}-{hc\over L^2},\quad -\delta b={hb\over H}-{hd\over L^2}
$$
Linear homogeneous system of equations connected $b,d$ lead as proper value
$\delta=h\sqrt {{1\over H^2}-{1\over M^2L^2}}$ and its solution in a form
$d={t\over M^2}, b=({1\over H}-{\delta\over h})t$
Non homogeneous system of equation for $a,c$ have the solution
$$
a={\delta\over h})t+{t_1\over L^2},\quad c=({1\over H}-{\delta\over h})t_1
$$  
After substitution these expressions into definition $\delta$ via them we come to relation connected parameters $t,t_1$ in a form
$$ 
{t^2\over M^2}+2({1\over H}-{\delta\over h})t t_1=h^2
$$
Checking the equations containing $\theta$ parameter show that all 3 equations are satisfy under the choice $\theta=h$.
The second Kazimir operator of (\ref{2}) looks as
$$
K_2=I^2+{(p)^2\over M^2}+{(x)^2\over L^2}-{(x p)+(p x)\over H}+
\sum_{i\leq j} (g_{i i}F_{i,j} g_{j,j}F_{i,j})({1\over H^2}-{1\over L^2M^2})\equiv 
$$
\begin{equation}
I^2+{(x-{L^2\over H}p)^2\over L^2}+({h^2\over H^2}-{h^2\over L^2M^2})
(-{L^2(p)^2\over h^2}+{\sum_{i\leq j} (g_{i i}F_{i,j} g_{j,j}F_{i,j})\over h^2})
\label{KAZ}
\end{equation}
From (\ref{2}) it follows that quadratical form in the last brackets is Kazimir operator of the group of 5 dimensional rotations in dimensionless form.

\section{Gauge field theory}

Let 10 generators of the group of the motion $p,f,l$ in dimensionless form denote with indexes generators of 5-th rotation group $P_{i,j}$. From (\ref{2
KAZ}) we have $P_{5,i}={L\over h}\bar p_i,P_{ij}={\theta\over h}(x_ip_j-x_jp_i)$ Then commutation relations between 
$P_{i,j}$ looks as
$$
[P_{i,j},P_{s,k}]=g_{j,s}P_{i,k}-g_{i,s}P_{j,k}-g_{j,k}P_{i,s}+g_{i,k}P_{j,s}
$$ 
After summation in both sides of this equation on indexes $s,j$ we obtain equality
$$
\sum_{j,s} g_{j,s}[P_{i,j},P_{s,k}]=3P_{i,k}
$$ 
Now let us construct Lie derivatives. To all operators of shifts in unphysical Minkovsky space add gauge potential. In other words in considered above realization $P_{i,j}$ via $p_i x_j$ exchange $p_i,\to p_i+ A_i,x_j\to x_j$, where as usually $A_i$ take values in some semi-simple algebra. Objects obtained in such way let us denote as $P^l_{i,j}$. Let us introduce   
$$
G_{i,k}=\sum_{j,s} g_{j,s}[P^l_{i,j},P^l_{s,k}]-3P^l_{i,k}, 1\leq i,j,..\leq 5
$$
Tensor of the field $G_{i,k}$ does not contain operators of differentiation on unphysical Minkovski space and under gauge transformation of potential $A_i\to SA_iS^{-1}+S_{x_i}S^{-1}$
transforms as $G_{i,k}\to SG_{i,k}S^{-1}$. 
Action  constructed with the help of Lagrangian function of gauge invariant theory with de Sitter group of motion looks as
$$ 
S=\int d^4 Y^{-4} Spur \sum_{i,j,k,s} g_{j,i}g_{k,s} G_{i,s} G_{j,k} 
$$
About calculation of invariant mesure see Appendix.
Now let us calculate all components of gauge field tensor. We regroup terms in $\bar p_i$ and represent it in equivalent form $\bar p_i=f(p_j+A_j)+b x_j(\rho-(x,p+A)$, where $f=a+b{1+x^2\over 2}$. From this moment index 5 will excluded and all latin litters take values of unphysical Minkowski space.
$$
G_{5,k}=\sum_s g^{j,s}[\bar p_j,G_{s,k}]=
$$
$$
\sum_s g^{j,s}[f(p_j+A_j)+b x_j(\rho-(x,p+A),x_s(p_k+A_k)-x_k(p_s+A_s)]=
$$
In further calculation only the following basic commutation relations are necessary
$$
F_{i,j}=[p_i+A_i,p_j+A_j],\quad [p_i+A_i,x_j]=g^{i,j},
$$
where now $F_{i,j}$ gauge field tensor of unphysical Minkovski space, in terms of which Lagrangian function will be represented.
Result of interrupted calculation is the following
$$
3\bar p_k+(f-b x^2)\sum_s g^{j,s}x_s F_{j,k}
$$
As was explained above the second term in the last equality is numerical value transforming as $\bar p_k$ (see Appendix) with respect to de-Sitter transformation and as tensor of the field with respect to gauge ones. Now
$$
\sum_s g^{A,B}[P^l_{i,A},P^l_{B,j}]=g^{5,5}[P^l_{i,5},P^l_{5,j}]+
\sum_s g^{s,s}[P^l_{i,s},P^l_{s,j}]
$$
Let us calculate first term with the same comments as was done few lines above 
$$
[f(p_i+A_i)+b x_l(\rho-(x,p+A)),f(p_j+A_j)+b x_j(\rho-(x,p+A))]=f^2[p_i+A_i,p_j+A_j]+
$$ 
$$ 
b(2f-b x^2)(x_i(p_j+A_j)-x_j(p_i+A_i))+f b(x_j \sum_s g^{s,s}x_s F_{s,i}-x_i 
\sum_s g^{s,s}x_s F_{s,j})
$$
By the same way calculation of the second term lead to
$$
\sum_{k,s} g_{k,s}[x_i(p_s+A_s)-x_s(p_i+A_i),x_k(p_j+A_j)-x_j(p_k+A_k)]=
$$
$$
2(x_i(p_s+A_s)-x_s(p_i+A_i))-x^2 F_{i,j}+x_i(x F)_j-x_j(x F)_i
$$
Keeping in mind condition of dimensionless in the first lines of this section summation of these two results lead to finally expression for $G_{i,j}$
$$
G_{i,j}=({L^2\over h^2}f^2-x^2)F_{i,j}-({L^2 \over h^2}f b-1)(x_i(x F)_j-x_j(x F)_i),\quad
G_{i,5}=-{L\over h}(f-b x^2) (x F)_i
$$
All coefficience in the expressions above may be expressed via one function $Y=(f-b x^2)=
a+{b\over 2}-{b x^2\over 2}$ as follows
$$
{L^2\over h^2}f^2-x^2={L^2\over h^2}Y^2,\quad {L^2 \over h^2}f b-1=-b{L^2 \over h^2}Y
$$
Now keeping in mind (\ref{KAZ}) and comments after it after simple manipulations we obtain density of Lagrangian function and corresponding action in a form
$$
L=Y^4 Spur \sum g_{ii}g_{jj}Spur(F_{i,j}F_{i,j}),\quad S=\int d^4 x Spur \sum g_{ii}g_{jj}Spur(F_{i,j}F_{i,j})
$$  
On the first look result is strange that we have action Lorenz invariant theory with translations. But it is necessary to keep in mind that theories with zero mass are invariant to additional conform transformation and thus invariant with respect to arbitrary linear combinations of translation and conform transformations. Generators $\bar p$ are exactly of this kind.  
And thus there are two possibilities construct theory with interaction with matter invariant to space time translation as it was done in the usual theory invariant to Poincare group,or construct the theory with interaction invariant with respect to de-Sitter group of five dimensional rotations. Exactly of this kind is unphysical Minkovski space considered in the present paper.  

The part of action responsible for interaction the gauge field with Dirac particles have the form
$$ 
S=\int d^4 xY^{-4} (\bar \psi (\sum \gamma_i \bar p^l_i+\sum (\gamma_i\gamma_j-
\gamma_j\gamma_i)\bar F^l_{i,j})\psi)
$$ 
where $\gamma_i,\bar \psi,\psi$ matrices of Dirac and  spinors functions ( four dimensional representation of de Sitter group). It is obvious that such constructed action is invariant as to gauge transformation as to transformation of de-Sitter group.

\section{Outlook}

The main result is the new alternative to Minkowski space-time picture of the universe.
We have challenged one of the fundamental principles of the usual theory - its invariance with respect to space-time translation. While the invariance under Lorenz transformations has been confirmed by all physics of the 20th century, the translation invariance has no direct experimental confirmation.
 
In the present paper we have rewritten results of \cite{LEZM} and generalized 
\cite{LEZp} in the form that is closest to the usual theory. To describe the world 
with 3 additional parameters it is necessary to solve equations of gauge field theory 
invariant with respect to de-Sitter group. The form of these equations are fixed uniquely from the conditions of gauge and de-Sitter invariance. Of course, first experimental results for or against of this theory should be expected from the cosmological data, where results of calculations will be in the greatest contradiction with the predictions of the usual theory. 

But in the physics of elementary particles the change should be no less revolutionary, 
because symmetry does not know anything about the distances, and replacing a  non-semi-simple Poincare algebra by a semi-simple de-Sitter algebra may lead to unexpected consequences. Moreover, the proposed theory of the Universe is not time invariant \cite{KHL} and this circumstance also may have unexpected repercussions both on microscopic and on macroscopic scales of observation.

Of course it is not possible exclude possibility that in future unphysical Minkowski space of the present paper become physical one. It was  many times in the science
when from un correct foundations was obtained new correct results.  We remind that result of Lorenz transformation was considered on duration of almost 30 years as unphysical ones. And only analyses of Einshtein return them into real world.

\section{Acknowledgments}

Author thanks J.Mostovoy for many fruitful discussions  and great help in preparation manuscript for publication and also CONNECUT for finance support.

\section{Appendix}

In this Appendix we would like to show how four dimensional manifold transformed under de-Sitter transformation (shifts $e^{(\alpha p)})$.
$x_i(t)=e^{(\alpha p)t}x_i e^{-(\alpha p)t}$ in connection with  results of section 3
$x_i(t)$ satisfy the following system of ordinary differential equations
$$
\dot {x_i}=(a+b{1+x^2\over 2})\alpha_i-b(\alpha x)x_i
$$
As a consequence we have equations for scalar function
$$
\dot {x^2}=(2a+b-b x^2)(\alpha x),\quad \dot {(\alpha x)}=(a+b{1+x^2\over 2})(\alpha)^2-
b(\alpha x)^2,
$$
$$
(2a+b-b x^2)\equiv Y,\quad (\alpha x)=-{1\over b}\dot {\log Y},\quad -{1\over b}\ddot {\log Y}=
(2a+b-{Y\over 2})(\alpha)^2-{1\over b}(\dot {\log Y})^2
$$
The last equation with the trivial manipulations takes the form
$$
\ddot {{1\over Y}}={(\alpha)^2h^2\over L^2}({1\over Y}-{b L^2\over 2h^2})
$$ 
with obvious solution
$$
{1\over Y}={b L^2\over 2h^2}+c_1 \sinh {(\alpha)h\over L}t+c_2 \cosh {(\alpha)h\over L}t
$$
Now the first equation above rewritten as
$$
\dot {{x_i\over Y}}=(2a+b-{Y\over 2})\alpha_i\equiv {\alpha_i\over (\alpha)^2b}\ddot {{1\over Y}}
$$
after this all other manipulations absolutely clear. 

With the results above it is possible to calculate invariant mesure of the space
$M(x^2) d^4 x$. Function $M$ satisfy the equation $ \dot {M}+(\sum_i (\dot {x_i})_{x_i})M=0$. Substituting all necessary values from the formulae above we obtain 
$M=cY^{-4}$. 

By the same arguments and technique we have for same transformation equations for $p$
$$
\dot {p_i}=-b x_i (\alpha p)+b\alpha_i [(x p)-\rho]+b(\alpha x)p_i,\quad \dot {(\alpha p)}={b(\alpha)^2\over 2}[(x p)-\rho],\quad {1\over Y}\dot{(x p)}={1\over 2}(\alpha p)+\dot {{1\over Y}}[(x p)-\rho] 
$$ 
After once differentiation the last equation and taking into account equation for ${1\over Y}$ we pass to equation for $[(x p)-\rho]$  
$$
\ddot {[(x p)-\rho]}={(\alpha)^2h^2\over L^2}[(x p)-\rho]
$$
Transformation above is similar to conformal invariance of zero interval in relativity
\cite{VAF}.

\end{document}